\documentclass[a4paper]{jpconf}
\usepackage{graphicx}
\begin{document}
\title{ Strange  Baryon to Meson Ratio}
\author{Eleazar Cuautle and Alejandro Ayala}
\address{Instituto de Ciencias Nucleares, Universidad Nacional
  Aut\'onoma de M\'exico, Apartado Postal 70-543, M\'exico Distrito Federal
  04510, M\'exico } 

\ead{ecuautle@nucleares.unam.mx}

\begin{abstract}
We present a model to compute baryon and meson transverse momentum
distributions, and their ratios, in relativistic heavy-ion
collisions. The model allows to compute the probability to form
colorless bound states of either two or three quarks as functions of
the evolving density during the collision. The qualitative differences
of the baryon to meson ratio for different collision energies and for
different particle species can be associated to the different density
dependent probabilities and to the combinatorial factors
which in turn depend on whether the quarks forming the bound states
are heavy or light. We compare to experimental data and show that we
obtain a good description up to intermediate values of $pt$.
\end{abstract}

\section{Introduction}

Long time ago it has been proposed~\cite{prl48-1982} that  strangeness is easier
produced in the QGP than in a hadronic environment. At the Large Hadron
Collider (LHC), new
results on strangeness production~\cite{exp-results} are being obtained, among them are the
transverse momentum distributions for pions, kaons, protons and
$\Lambda$ as well as baryon to meson ratios. The
distributions show disagreement with perturbative QCD 
results, especially on the strangeness sector. Consequently, the
baryon to meson ratio presents an unexplained behavior at intermediate
transverse momenta.
\noindent
The  baryon to meson ratios measured at LHC energies could be
different from the ratios measured at lower energies, since the former are driven predominantly by hard collisions.
Many ideas have been developed to
understand these ratios. Some  studies include the percolation model~\cite{prc82-2010},
recombination~\cite{prc84-2011+}, the statistical model~\cite{CPC180-2009+}, and also flow~\cite{jpg-prl111-2013} in proton-proton collisions.
The slight discrepancies of the statistical model with data provide the motivation to get further insight into the hadronization mechanism~\cite{prc74-2006}. 

This  work reports on an alternative model to study the production of hadrons
in relativistic heavy ion collisions, where the evolving density during the collision and the strangeness abundance are key ingredients, the so called {\it dynamical quark recombination model}~\cite{prc77-2008+}. 

\section{Transverse momentum distribution }

The transverse momentum
distributions are computed within the dynamical recombination
model are given by 

\begin{eqnarray}
   \frac{dN}{p_T dp_T dy}&=&
   g\frac{m_T}{4\pi}\frac{\rho_{\mbox{\tiny{nucl}}}^2}{\Delta\tau}
   \int_{\tau_{0}}^{\tau_{f}}\tau d\tau{\mathcal{P}}(\tau )
   I_0(p_T\sinh\eta_T / T) \nonumber \\
   & \times&
   \int d\eta\cosh(y - \eta )e^{-[m_T\cosh(y - \eta )\cosh\eta_T]/T} 
   \label{distfin}
\end{eqnarray}

\noindent
where $\Delta \tau=\tau_f-\tau_0$ is the proper-time interval spanned during the hadron emission, $\eta_T$ is related to the transverse expansion velocity $\beta_T$ by $\beta_T =\tanh(\eta_T)$,
the temperature $T$ is given by the cooling law $ T=T_0 (\tau_0/\tau)^{v_s^2}$ and 
${\mathcal{P}}$ is the probability of forming a given hadron as a function of proper-time $\tau$ and accounts for the fact that hadronization is not instantaneous. The  profile
 of this probability can be obtained by Monte Carlo simulation (see
 Ref~\cite{prc77-2008+} for details).
The spectrum  given by Eq.~(\ref{distfin}), has $T$, $\beta_T$ and the overall normalization as the free
parameters.

\subsection{Combinatorial probabilities}

The combinatorial analysis allows to estimate in a simple way the number of possible
colorless baryons and mesons, when we take a certain number of heavy ($Q$)
and light ($q$) quarks.  Considering the case where one starts out with
a set of $n$ $q$-quarks ($q = u,d$) and $m$ $Q$-quarks ($Q=s$), it is
possible to form sets of two (mesons) or tree (baryons) quarks. The
more simple case is when we have one $q$ and one $Q$ each comming in
three colors.  In that case the possible colorless combinations are
show on the table~\ref{table1}. From this it is easy to compute   the
relative abundance of the mesons containing a $Q$ quark with respect to the
total combinations, so the  number of relative mesons is  $2mn/(n+m)^{2}$. In a similar way, the  total number of  possible baryons
plus antibaryons is $2(n+m)^{3}$, and the relative abundances of the
baryons containing one $Q$ quark is  $3n^{2}m/(n+m)^{3}$.
In this way,  one can compute  the baryon to meson ratio as a function of the
number of light ($n$) and heavy ($m$) quarks.


\begin{center}
\begin{table}[h]
\caption{\label{table1}Number of possible combination of 2 or 3 quarks }
\centering
\begin{tabular}{@{}*{7}{ll}}
\br
Kind & \# of Mesons & &  kind & \# of Baryons\\
\mr
$q \bar q$ &  $3n^{2}$ & & $qqq$ & $n^{3}$\\
$q \bar{Q}$ & $3nm$   & & $qqQ$  & $3n^{2}m$  \\
$Q \bar{q}$ & $3nm$   & & $qQQ$  & $3nm^{2}$  \\
$Q \bar{Q}$ & $3m^{2}$ & & $QQQ$  & $m^{3}$  \\
\br
\end{tabular}
\end{table}
\end{center}

\noindent
The previous exercise may  be generalized for more light  flavors, for
instance, for the case of $q= u, d$ and $Q=s$. In this case, some  of the
combinations are  $\Omega(sss)$ and $\Phi (s\bar s)$.
Considering an equal number ($n$) of $u$ and $d$ and $m$ as
the number of $s$-quarks, we can estimate the ratio $\Omega/\Phi$ as,

\begin{equation}
\frac{\Omega}{\phi} = \frac{4m^2 n+m^2}{8n^3 + 3n^2 m +6 n m^2}.
\label{OmegaPhi-nm}
\end{equation}

\noindent
Since the strange quarks are produced in lesser quantities than up and down, we introduce the variable $x > 1$ as $ u = x s$ to write Eq.~(\ref{OmegaPhi-nm}) as

\begin{equation}
\frac{\Omega}{\phi} = \frac{4 x +1}{8x^3+ 3x^2 +6x}.
\label{OmegaPhi-ud-xs}
\end{equation}

\noindent
This equation could  be simplified even more, when ones takes only one
light and one heavy flavor. Let us take for instance only  $u$ and $s$
quarks, then Eq.~(\ref{OmegaPhi-ud-xs}) becomes 

\begin{equation}
\frac{\Omega}{\phi} = \frac{1}{x+1}.
\label{OmegaPhi-d-xs}
\end{equation}

\begin{figure}[h]
\begin{minipage}{14pc}
\includegraphics[width=20pc]{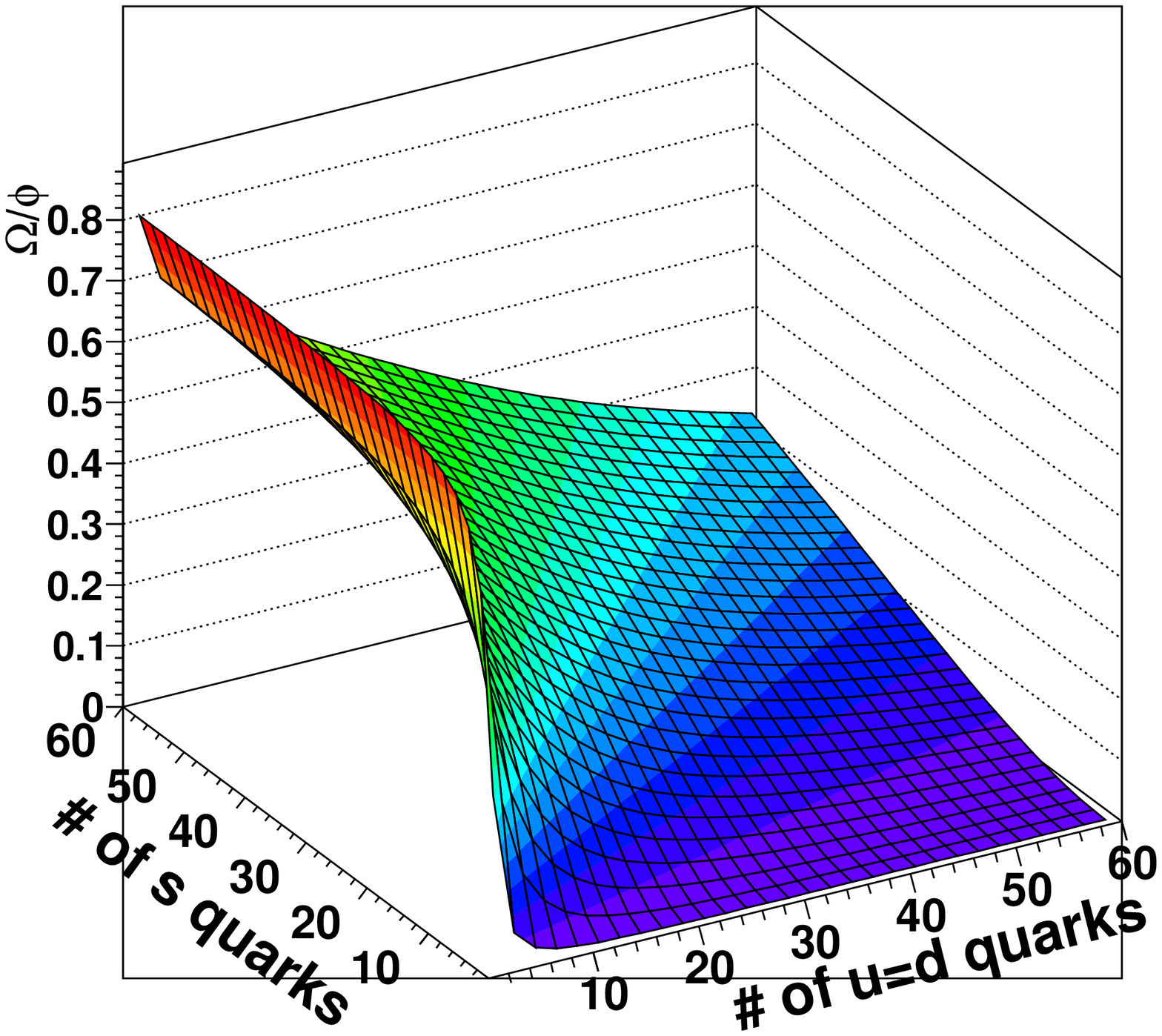}\hspace{1pc}%
\caption{\label{fig1}Ratio $\Omega/\Phi$ as function of the number of
  $s$ and light ($u=d$) quarks.}
\end{minipage}\hspace{6pc}%
\begin{minipage}{14pc}
\includegraphics[width=20pc]{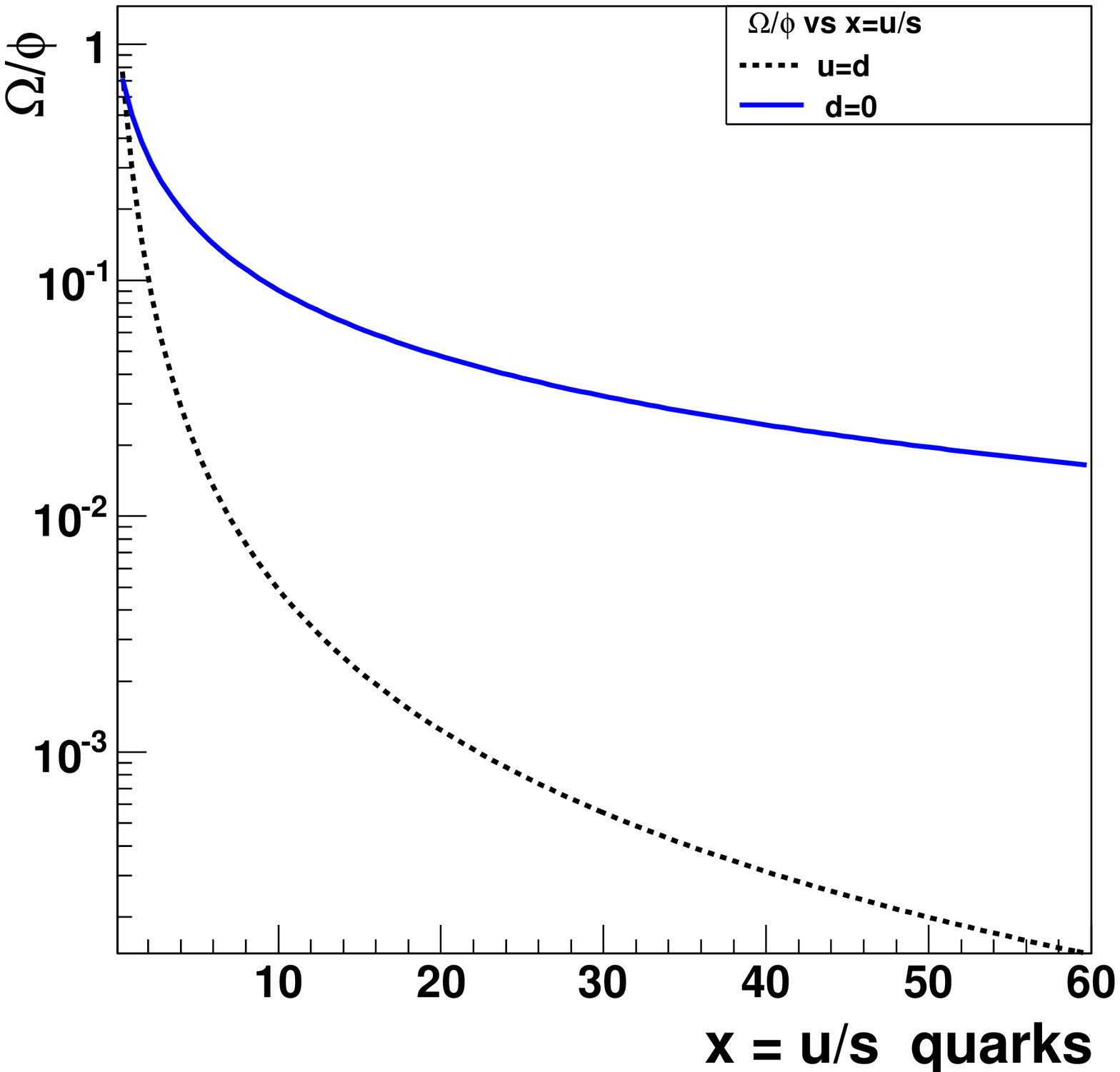}\hspace{1pc}%
\caption{\label{fig2} Ratio $\Omega/\phi$  as function of
  proportionality factor $x=u/s$ between light $u,d$ and $s$ quarks.}
\end{minipage}
\end{figure}

\noindent
The behavior of the $\Omega/\Phi$ ratio   as a function of the number of
light and heavy flavors, described by Eq.~(\ref{OmegaPhi-nm}), is shown
in Fig.~\ref{fig1}.  For the case when we take $u =x s$ and the number of $u$'s is equal to the number
of $d$'s,  the baryon to
meson ratio has the behavior shown with the dashed curve in Fig.~\ref{fig2}. The case where there is  only one heavy and one light quark ($d=0$), described by Eq.~(\ref{OmegaPhi-d-xs}),
is shown in the same figure with the solid curve. Notice that low values of $x$ correspond to higher collisions energies whereas high values correspond to less energetic collisions.

\section{Results}

\begin{figure}[h]
\begin{minipage}{14pc}
\includegraphics[width=20pc]{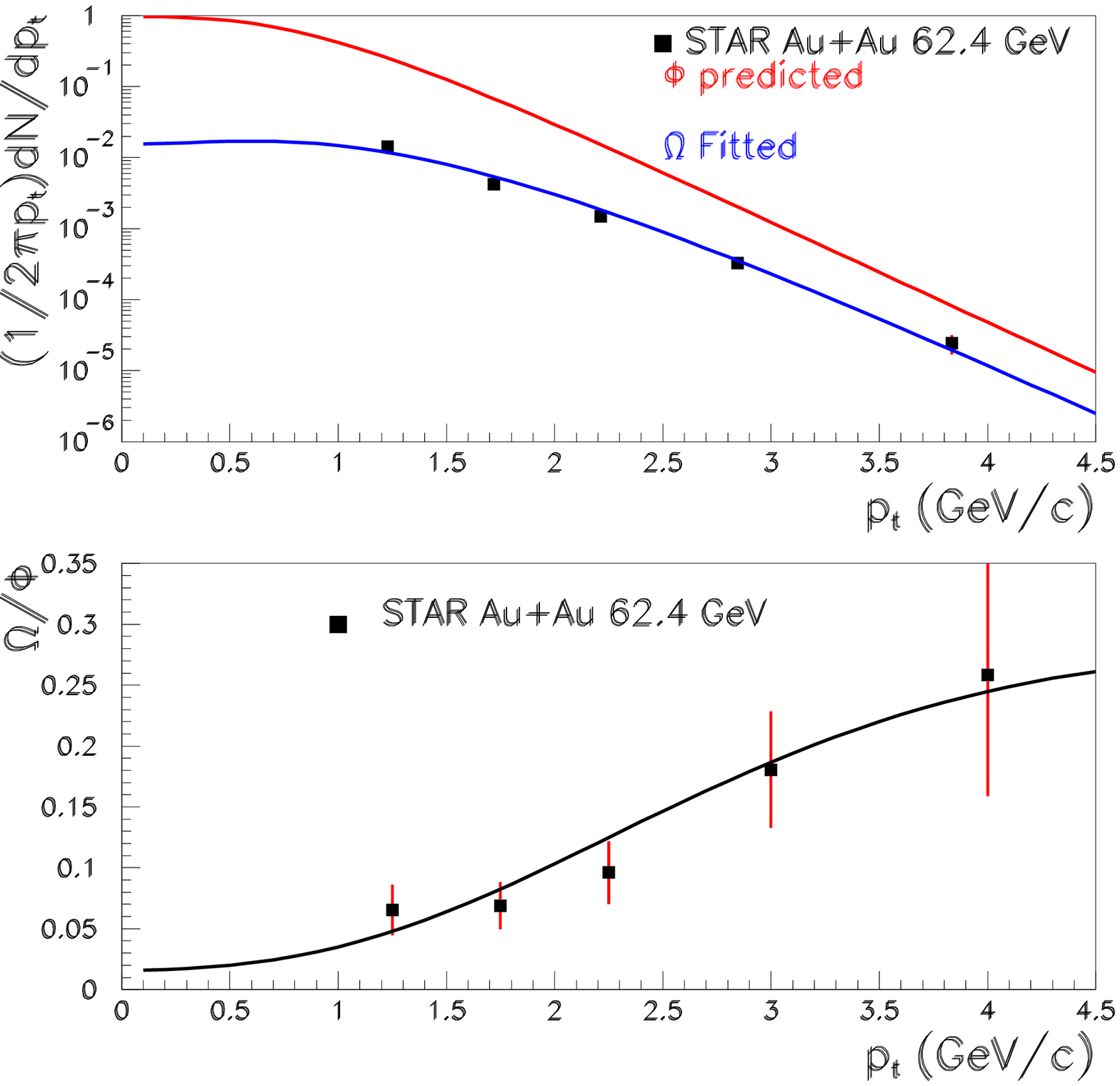}\hspace{1pc}%
\caption{\label{fig3}Experimental results of the spectra and the ratio
  $\Omega/\Phi$ fitted to the model and extracted the $\beta =0.40$} 
\end{minipage}\hspace{6pc}%
\begin{minipage}{14pc}
\includegraphics[width=20pc]{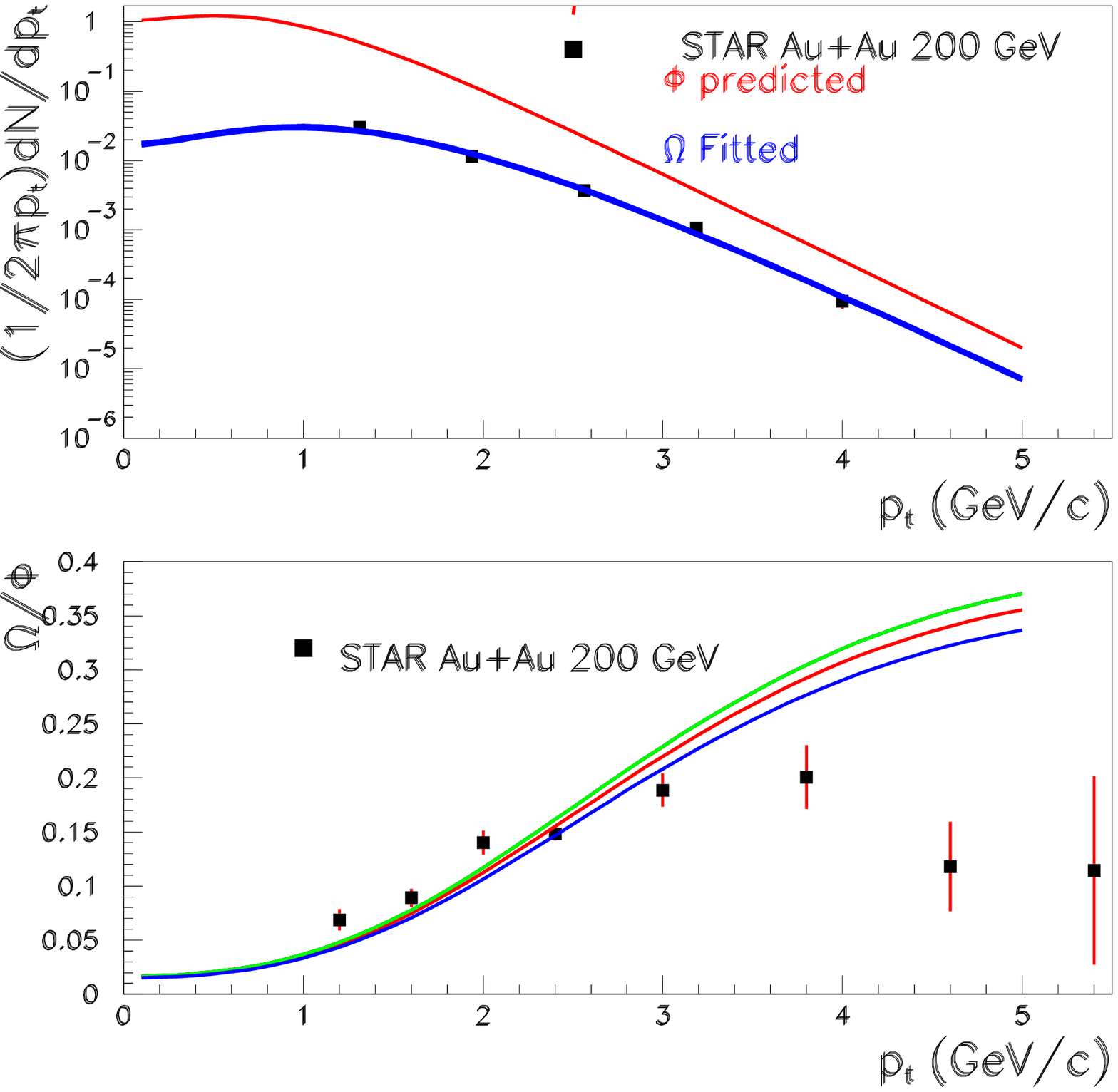}\hspace{1pc}%
\caption{\label{fig4}
Spectra and ratio $\Omega/\phi$  fitted to the model and extracted the
$\beta =0.45$, and the ratio for three values of $x$.}
\end{minipage}
\end{figure}

In order to estimate the momentum distribution, we take an initial
hadronization time, $\tau_{0} = 1$ fm, and initial and final temperatures as
200 and 100 MeV respectively, which means a final
hadronization time $\tau_{f} =8$ fm. Taking the  $\Omega$ spectrum
produced in  Au+Au from STAR data~\cite{STAR-data-62Gev} at 62.4 GeV, we can fit it
and extract the transverse velocity $\beta_{T}$, as well as the normalization
constant. The last parameters and the data on the ratio $\Omega/\Phi$
are used to extract the distribution of the $\Phi$, since this last one
is not reported from experiment. The results are plotted in Fig.~\ref{fig3}. 
Regarding the combinatorial  probability and $p_t$ spectra, let us
take into account the constrains on the probability ratios to form  baryons  and mesons,  and
the  the fact that there is  a proportionality factor $x$ between light and heavy quarks. Furthermore,
taking into account the parameters of the fit and the constrains of
the Eq.~(\ref{OmegaPhi-d-xs}), it is possible to extract the
proportionality factor $x$ between the number of light and heavy
quarks. The top of Fig.~\ref{fig4} shows, the fitted spectra to
$\Omega$ data and the  prediction of the $\Phi$ spectrum. The bottom part shows the ratio fitted with
three different values of the $x$. One can see that after $\approx$ 3 GeV our curves grow above data. This behavior could be explained  as a consequence of the absence of the fragmentation and energy loss in our approach.

In summary, the dynamical recombination model together with
constrains on the $\Omega/\phi$ ratio from combinatorial probabilities can describe RHIC data for Au+Au at two energies.  We observe that the fits are better
at low energy and lighter ions in the collisions. The constrains on
the probabilities could provide a reasonable value for the
proportionality between the number of light and heavy flavors, $x$. 

\subsection{Acknowledgments}

Support for this work has been received by CONACyT
under  grant numbers 101597 and 128534
and PAPIIT-UNAM under grant numbers IN107911 and IN103811.


\end{document}